# Advanced Health Misinformation Detection Through Hybrid CNN-LSTM Models Informed by the Elaboration Likelihood Model (ELM)

Mkululi SIKOSANA
*Department Of Computing And Mathematics*
Manchester Metropolitan University
Manchester, UK
mkululi.sikosana@stu.mmu.ac.uk

Sean MAUDSLEY-BARTON
*Department Of Computing And Mathematics*
Manchester Metropolitan University
Manchester, UK
s.maudsley-barton@mmu.ac.uk

Oluwaseun AJAO
*Department Of Computing And Mathematics*
Manchester Metropolitan University
Manchester, UK
s.ajao@mmu.ac.uk

*Abstract*—Health misinformation during the COVID-19 pandemic has significantly challenged public health efforts globally. This study applies the Elaboration Likelihood Model (ELM) to enhance misinformation detection on social media using a hybrid Convolutional Neural Network (CNN) and Long Short-Term Memory (LSTM) model. The model aims to enhance the detection accuracy and reliability of misinformation classification by integrating ELM-based features such as text readability, sentiment polarity, and heuristic cues (e.g., punctuation frequency). The enhanced model achieved an accuracy of 97.37%, precision of 96.88%, recall of 98.50%, F1-score of 97.41%, and ROC-AUC of 99.50%. A combined model incorporating feature engineering further improved performance, achieving a precision of 98.88%, recall of 99.80%, F1-score of 99.41%, and ROC-AUC of 99.80%. These findings highlight the value of ELM features in improving detection performance, offering valuable contextual information. This study demonstrates the practical application of psychological theories in developing advanced machine learning algorithms to address health misinformation effectively.

## I. Introduction

The COVID-19 pandemic has challenged global health systems and highlighted the rapid spread of health misinformation on social media platforms. Health misinformation can lead to harmful consequences, including vaccine hesitancy, the spread of unfounded medical treatments, and increased public anxiety [1][2]. Consequently, detecting and mitigating misinformation has become a critical priority for public health professionals and associated agencies [3]. Despite significant advancements in computational methods, network analysis, linguistic analysis, and data-driven approaches to misinformation detection [4][5][6][7], the application of psychological theories remains limited. For this reason, this study attempts to apply the theoretical framework of the ELM [8] to inform COVID-19 health misinformation detection using a Twitter dataset. Specifically, our study aims to (1) examine message characteristics that influence the likelihood of health misinformation and (2) identify the central and peripheral routes informing health misinformation detection. In this work, we focus on how health misinformation discourse in online social networks (OSN) is created, spread, and discussed across various social media platforms [9]. In doing so, we consider multiple dimensions, including the nature of the misinformation, the platform used for dissemination, the message characteristics, the role of algorithms, and the impact on public health [10][11]. We explore the uses of ELM to enhance current algorithms for health misinformation. While several studies have explored misinformation detection (e.g., [6][9]), the specific application of ELM to health misinformation detection using hybrid DL appears to be relatively unexplored in the current literature. While some studies have examined ELM in online health contexts (e.g., [33][36][37]), its use in automated health misinformation detection with hybrid DL models remains largely novel. We hypothesise that integrating ELM-based features into the machine learning (ML) model will enable the enhanced model to outperform the base model in accuracy, precision, recall, F1-score, and ROC-AUC. To our knowledge, this is the first study to examine the effectiveness of ELM in the context of health misinformation detection on social media. The findings are expected to inform the development of robust algorithms that public health professionals can use to detect and mitigate health misinformation in OSN, thereby supporting appropriate strategies to counteract misinformation.

This study contributes to the field of health misinformation detection on OSN by:

- Integrating the ELM with ML techniques, mainly through developing a hybrid CNN-LSTM model. This integration leverages cognitive and affective features derived from ELM (such as sentiment polarity, text length, and readability) to enhance misinformation detection.
- Employing advanced natural language processing (NLP) for automated feature extraction.

## II. Related work.

A plethora of literature explores various aspects of health misinformation on social networks [6][9] alongside related research areas. The following review examines key themes across the literature to identify common threads, gaps, and opportunities and to demonstrate how our study addresses these gaps by focusing on a Twitter dataset and a hybrid CNN-LSTM model.

*A. Role of Sentiment and Emotional Appeals*

[14] and [15] highlight the importance of emotional appeals in disseminating health misinformation. [14] used ELM and congruence theory in analysing COVID-19 debunking posts, revealing a significant impact of both cognitive and affective appeals on message persuasiveness. However, their reliance on a single platform (Sina Weibo, often called the Chinese Twitter) limits the generalisability of their findings to other social media. Furthermore, the study's strong explanatory power ($R^2=0.658$) indicates robust findings, yet the methodological approach of using ridge regression could be complemented with more diverse ML techniques to validate results across different datasets. [15] expanded on this approach by examining the retweet likelihood of pro- and anti-COVID-19 vaccine messages on Twitter. Their use of logistic and generalised negative binomial regressions offers insights into the differential impact of content-related and content-unrelated characteristics on message dissemination. While that study's findings are statistically significant, the heavy reliance on regression models may overlook more nuanced patterns that advanced ML models could detect [34]. [16] underscores the influence of sentiment on information dissemination on Weibo, with their sentiment analysis using BERT achieving 76% accuracy. The study's regression models show that negative sentiment is significantly associated with increased forwarding volume, spreading depth, and network influence (β values 0.029–0.034, $p<0.001$). Nonetheless, the study could benefit from exploring interactions between sentiment and other textual features, such as readability and text length, to provide a more comprehensive understanding of misinformation propagation.

[17] and [21] focus on the continuous usage intentions of mHealth services and the credibility of online consumer reviews, respectively. Both studies utilise structural equation modelling (SEM) to reveal significant determinants of user behaviour, with [17] achieving $R^2=0.67$ for continuous usage intention. While these findings are insightful, they lack direct application to misinformation detection, necessitating further research to adapt these models for misinformation contexts.

*B. Central and Peripheral Route Processing*

[33] and [14] effectively utilise the ELM to explore central and peripheral route processing in contexts such as vaccination intentions and debunking messages. [33] find that peripheral-route factors such as emotional contagion and media credibility have a more substantial impact on panic vaccination intention, with their partial least squares-SEM explaining 55.8% of the variance. However, the study's focus on survey responses limits its applicability to real-time social media data. [14] combined linguistic analysis with ML to analyse COVID-19 debunking posts, emphasising the need to incorporate cognitive (central) and affective (peripheral) appeals into automated misinformation detection models. Their approach is innovative, but it could be enhanced by integrating temporal dynamics to capture the evolving nature of misinformation. [34] provides a comprehensive framework for fake news detection using a multi-cue approach (achieving over 80% accuracy and ~90% recall); while robust, that study could benefit from validating findings across multiple social platforms and exploring the interplay between different types of cues. [35] investigated the influence of website design elements on online persuasion, finding significant effects of argument quality and design elements on user attitudes (their PLS-SEM model explains 35.7% of the variance in issue involvement change). However, because [35] focuses on website design, its direct applicability to social media misinformation detection is limited. [36] and [37] explore the perceived credibility of rebuttals and belief in false news through the ELM framework. Both studies highlight the importance of information quality and source credibility, with [36]'s model explaining a substantial portion of the variance in perceived credibility. These findings underscore the need to incorporate credibility-oriented factors into misinformation detection models. [20] demonstrates the importance of argument quality and source credibility in information dissemination behaviour (ELM-based, $R^2=0.59$ for sharing intentions). Like many others, however, that study relies on survey data, which may not fully capture the complexities of real-world misinformation dynamics. The heavy use of survey and experimental methods in these works limits their real-time applicability to social media contexts.

*C. Platform-Specific Analyses*

[15] focuses exclusively on Twitter, identifying how specific message characteristics influence retweet behaviour. Conversely, [16] and [33] examine Sina Weibo (the Chinese context), highlighting the impact of sentiment and other factors on information dissemination. These platform-specific analyses provide valuable insights but limit the generalisability of findings across different social media platforms. [29] examines the relationship between news exposure and trust in COVID-19 fake news on social media, finding that increased exposure to credible news sources decreases the likelihood of believing fake news. However, [29] primarily relies on regression analysis, which may not capture the nuanced patterns of misinformation spread across platforms. Cross-platform studies are needed to validate these insights and enhance generalisability.

*D. Integration of Cognitive and Affective Cues with Machine Learning*

[38] explored persuasive strategies in COVID-19 vaccine-related Twitter posts, integrating ELM with Social Judgment Theory (SJT) and the Extended Parallel Process Model (EPPM). Their Random Forest (RF) classifier achieved an overall accuracy of 74.6%, highlighting the effectiveness of combining multiple theoretical perspectives with ML techniques. However, there is a need to investigate how such cues can be incorporated into hybrid DL models (e.g., combining CNN and LSTM architectures) for more robust misinformation detection. [32] demonstrated the potential of integrating textual analysis and ML for predictive modelling in the context of crowdfunding, achieving 73% accuracy (71% F-measure), an improvement

over baseline models. This approach suggests that incorporating diverse textual features and advanced ML techniques can improve predictive performance, a strategy that can also be adapted to enhance misinformation detection models. Overall, while these studies demonstrate the value of integrating cognitive and affective cues, developing hybrid models (e.g., CNN-LSTM combinations) remains an opportunity for achieving more robust performance in misinformation classification.

*E. Theoretical and Methodological Foundation for Our Approach*

Our literature review revealed several critical gaps that informed our methodological approach. First, while sentiment analysis has been widely employed in misinformation detection (as seen in many of the works above), studies have typically treated sentiment as an isolated feature rather than integrating it within a comprehensive psychological framework [40]. The ELM offers a structured approach to organising features into meaningful cognitive pathways, combining sentiment with other message characteristics. Moreover, prior works often rely on simple models (e.g., regression), highlighting the need for more complex integrations. Second, existing DL approaches frequently employ either convolutional networks for spatial text feature extraction or recurrent networks for sequential patterns but rarely combine these architectures to capture both dimensions simultaneously, leaving the potential of hybrid models underexplored [4]. Third, preliminary analysis of our COVID19-FNIR dataset revealed distinct patterns in content features (aligned with central-route processing) and stylistic elements (aligned with peripheral-route processing), suggesting that a dual-processing approach would be most effective. To validate this approach, we conducted an exploratory data analysis on a subset of 500 tweets, finding significant differences in readability metrics ($p < 0.01$) and punctuation patterns ($p < 0.05$) between misinformation and legitimate health information. These findings provided empirical support for pursuing the ELM-informed hybrid model described in the methodology.

**Formal Problem Definition**

Let $D = \{(x_1, y_1), (x_2, y_2), ..., (x_n, y_n)\}$ represent the dataset where $x_i$ is the i-th tweet text and $y_i \in \{0, 1\}$ is the corresponding label (0 for authentic news, 1 for fake news).

The classification task can be defined as finding the optimal function f: $X \rightarrow Y$ that maps the input space X to the output space $Y = \{0, 1\}$ by minimizing a loss function $\mathcal{L}$.

**ELM Feature Formulation**
**(i) Central Route Features**

Let $C(x_i)$ represent the set of central route features for tweet $x_i$:
$C(x_i) = [c_1(x_i), c_2(x_i), ..., c_m(x_i)]$
where:
$c_1(x_i)$ = Flesch-Kincaid Grade Level =
$$0.39 \left(\frac{total\ words}{total\ sentences}\right) + 11.8 \left(\frac{total\ syllables}{total\ words}\right) - 15.59$$

$c_2(x_i)$ = Vocabulary Richness
$$\frac{|unique\ tokens\ in\ x_i|}{|total\ tokens\ in\ x_i|}$$

$c_3(x_i)$ = Sentiment Polarity =
$$\frac{\sum_{j=1}^{|x_i|} polarity(w_j)}{|x_i|}$$
where polarity($w_j$) is the sentiment score of word $w_j$

$c_4(x_i)$ = Text Length = $|x_i|$ (number of tokens)

$c_5(x_i)$ = Average Words Per Sentence =
$$\frac{total\ words\ in\ X_i}{total\ sentences\ in\ X_i}$$

**(ii) Peripheral Route Features**
Let $P(x_i)$ represent the set of peripheral route features for tweet $x_i$:
$P(x_i) = [p_1(x_i), p_2(x_i), ..., p_k(x_i)]$

where,
$p_1(x_i)$ = Exclamation Mark Ratio =
$$\frac{count\ of\ "!\ "\ in\ x_i}{|x_i|}$$

$p_2(x_i)$ = Question Mark Ratio =
$$\frac{count\ of\ "?"\ in\ x_i}{|x_i|}$$

$p_3(x_i)$ = Capitalization Ratio =
$$\frac{count\ of\ upper\ case\ words\ in\ x_i}{total\ words\ in\ x_i}$$

$p_4(x_i)$ = All-caps Words Count =
$|\{w_j \in x_i : w_j\ is\ all\ uppercase\}|$

$p_5(x_i)$ = Urgency Terms Frequency =
$$\frac{|\{w_j \in x_i : w_j \in urgency\ lexicon\}|}{|x_i|}$$

**(iii) Combined ELM Features**
The complete ELM feature vector $E(x_i)$ for tweet $x_i$ is the concatenation of central and peripheral features:
$E(x_i) = [C(x_i), P(x_i)]$

### III. MATERIALS AND METHODS.
*A. Dataset and Preprocessing*

We evaluated our approach on the COVID-19 Fake News Infodemic Research Dataset (COVID19-FNIR) [39]. This dataset is a class-balanced collection of 7,588 news items, equally distributed into true and fake news classes (approximately 50% true, 50% fake). The data were sourced from Poynter (for fake news) and verified news publishers' Twitter accounts (for true news). Each entry contains attributes such as Text, Date, Region, Country, Explanation, Origin, and Label, though our study focuses solely on the textual content and its label (true or fake). The dataset is provided in two files (trueNews.csv and fakeNews.csv) with 8 and 12 columns, respectively; notably, true news texts often include links to additional information, unlike the fake news texts. To ensure uniformity, extensive text preprocessing was applied. In particular, all text was lowercased and cleaned by removing URLs, special characters, and excess whitespace to standardise the inputs. We then combined the data and employed a stratified 10-fold cross-validation approach (preserving the 1:1 class ratio in each fold) to robustly train and evaluate our models.

*B. Feature Engineering Based on ELM*

Following the ELM framework, we extracted two categories of features from each news text: features reflecting the *central route* (which involve deeper cognitive processing of message content) and features reflecting the *peripheral route* (which involves superficial cues and heuristics).

*1)* **Central Route Features:** These features capture aspects of the text that would be scrutinised under high elaboration. We computed several readability metrics, including the Flesch–Kincaid Grade Level, to gauge the complexity of the text. We also measured vocabulary richness (a lexical diversity measure, e.g., the ratio of unique tokens to total tokens) to indicate the content level of detail or variety. In addition, we performed sentiment analysis on each text using the TextBlob library, obtaining a sentiment polarity score that reflects the message's overall emotional tone (from negative to positive). Higher readability (indicative of more complex content) and sentiment polarity are central-route factors, as readers engaging with the content critically would process these aspects. We also note features such as the length of the text (e.g., total word count) and average words per sentence, which contribute to how much substance the message contains.

*2)* **Peripheral Route Features**: These features represent surface-level cues that might influence readers' acceptance of the message without deep analysis. We extracted punctuation usage patterns, such as the count of exclamation points and question marks, since messages with excessive punctuation can serve as peripheral cues (e.g., conveying urgency or emphasis). We also considered the presence of simple lexical cues, such as common n-grams or all-caps words, that could act as heuristics. For example, the frequency of specific bigrams/trigrams or capitalised words might catch attention and influence perception via the peripheral route. Additionally, although the dataset did not include explicit source information for each tweet, we acknowledge that in a broader context, source credibility indicators (such as verified account status or follower count) are important peripheral cues; however, our feature set is limited to textual characteristics available in the data.

*C. Hybrid CNN-LSTM Model Architecture and Training*

DL experimental setup for CNN-LSTM models. Our detection approach uses a hybrid CNN-LSTM neural network implemented in Keras to combine the benefits of convolutional and recurrent architectures. We first transform each tweet's text into a sequence of word embeddings. We utilised Keras' Embedding layer to vectorise words (with an embedding dimension of 100) based on the corpus vocabulary. The CNN component consists of a 1D convolutional layer with 64 filters (kernel size 3) that extracts local textual patterns (e.g., key phrases or n-grams) from the embeddings. The resulting feature maps are fed into a dropout layer (dropout rate 0.5) to reduce overfitting. Next, an LSTM layer with 100 units processes the sequence of CNN-derived features to capture the text's long-term dependencies and contextual information. The LSTM output provides a rich representation of the tweet's content. In the enhanced model, this text-derived representation is concatenated with the additional ELM features described in Section 3.2 (after scaling them to a comparable range). This combined feature vector (textual + engineered features) is passed through a dense layer with sigmoid activation to perform binary classification (fake vs. true). This architecture allows the model to leverage detailed content understanding and peripheral cues. We trained the models using the binary cross-entropy loss function and the Adam optimiser (learning rate 0.001). Each model was trained for 10 epochs with a batch size of 32, using early stopping on validation loss to prevent overfitting. The base model was trained using only the text (CNN-LSTM) inputs, whereas the enhanced model was trained with both text and additional ELM features. All models were evaluated using the cross-validation procedure described above. It should be noted that no external pre-trained embeddings or transfer learning were used; all features were learned from the training data.

The experimental setup evaluated the hybrid CNN-LSTM with Keras embeddings on the COVID19-FNIR DATASET for fake news detection, and we used the Keras embedding technique to preprocess text data from the COVID19-FNIR DATASET. The CNN layer, with 64 filters, extracted local textual features, which were then fed into an LSTM layer with an output dimension of 100 to capture long-term dependencies. A dropout layer reduced overfitting, and the output layer, with a sigmoid activation function, classified text as fake or true. This setup integrated CNN's feature extraction with LSTM's sequential data processing, providing a comprehensive approach to fake news detection in the COVID-19 context. The ELM feature extraction involved utilising central and peripheral routes to extract features such as 'flesch_kincaid_grade', 'vocabulary_richness', and 'sentiment_polarity' using the TextBlob library for sentiment and readability analysis. TextBlob processes the text to calculate sentiment strength and polarity, while other features such as 'text_length' and 'avg_words_per_sentence' are derived from basic text statistics. This comprehensive feature set captures detailed argumentative content and superficial stylistic cues for enhanced fake news detection. All models share the architecture described above; the enhanced model includes additional inputs from ELM-based features. After extracting central and peripheral cues using the ELM framework, we integrate these into a hybrid CNN–LSTM pipeline described below, to capture semantic and sequential patterns for robust misinformation classification.

**CNN Component**
**Word Embedding Layer**

Let $W \in \mathbb{R}^{|V| \times d}$ be the word embedding matrix, where $|V|$ is the vocabulary size and $d$ is the embedding dimension. For each token $w_j$ in tweet $x_i$, we obtain its embedding vector $e_j \in \mathbb{R}^d$.

The sequence of word embeddings for tweet $x_i$ is represented as:

$E_i = [e_1, e_2, ..., e_{(x_i)}] \in \mathbb{R}^{|x_i| \times d}$

**Convolutional Layer**

For a kernel size $h$, the convolutional operation applies a filter $F \in \mathbb{R}^{h \times d}$ to a window of $h$ words to produce a new feature. For the $j$-th window in tweet $x_i$, the feature $c_j$ is generated as:

$c_j = \text{ReLU}(F \cdot E_i[j:j+h-1] + b)$

where,
$b$ is the bias term and ReLU is the activation function.
With $nf$ filters, we obtain a feature map:
$C_i \in \mathbb{R}^{(|x_i|-h+1) \times nf}$ for tweet $x_i$.

**Max-Pooling Operation**

The max-pooling operation extracts the most important feature from each filter's feature map:
$mf = \max_j [c_{(j,f)}]$
where $c_{(j,f)}$ is the $j$-th feature generated by the $f$-th filter.
This results in a pooled feature vector $M_i = [m_1, m_2, ..., m_{(nf)}] \in \mathbb{R}^{nf}$ for tweet $x_i$.

**LSTM Component and Formulation**

The LSTM processes the CNN feature maps $C_i$ sequentially. For each time step $t$, the LSTM computes:
$i_t = \sigma(W_{(ii)} \cdot C_{(i,t)} + W_{(hi)} \cdot h_{(t-1)} + b_i)$
$f_t = \sigma(W_{(if)} \cdot C_{(i,t)} + W_{(hf)} \cdot h_{(t-1)} + bf)$
$g_t = \tanh(W_{(ig)} \cdot C_{(i,t)} + W_{(hg)} \cdot h_{(t-1)} + bg)$
$o_t = \sigma(W_{(io)} \cdot C_{(i,t)} + W_{(ho)} \cdot h_{(t-1)} + bo)$
$c_t = f_t \odot c_{(t-1)} + i_t \odot g_t$
$h_t = o_t \odot \tanh(c_t)$

where:
• $i_t, f_t, o_t$ are the input, forget, and output gates
• $g_t$ is the cell input activation
• $c_t$ is the cell state
• $h_t$ is the hidden state
• W terms are weight matrices
• b terms are bias vectors
• $\sigma$ is the sigmoid function
• $\odot$ represents element-wise multiplication

The final hidden state $h_{(|x_i|)}$ contains the sequence representation.

**Enhanced Model with ELM Features**

The enhanced model combines the LSTM output with the ELM features:
$z_i = [h_{(|x_i|)}, E(x_i)]$

This combined representation is then passed through a fully connected layer to obtain the classification probability:
$\hat{y}_i = \sigma(W_z \cdot z_i + b_z)$
where $W_z$ is the weight matrix and $b_z$ is the bias vector for the fully connected layer.

*D. Evaluation Metrics and Validation*

We evaluated model performance using standard metrics after training and testing across 10 cross-validation folds, reporting aggregate results. The metrics include accuracy, precision, recall, F1-score, and ROC-AUC. Accuracy measures overall correctness, calculated as (TP + TN) / (TP + TN + FP + FN), where TP (true positives) are correctly identified fake instances, TN (true negatives) are accurately recognized true instances, FP (false positives) are true instances misidentified as fake, and FN (false negatives) are missed fake instances. Precision indicates the proportion of actual fake identifications, calculated as TP / (TP + FP), reflecting the model's predictive accuracy. Recall (sensitivity) measures the proportion of actual fake news correctly identified, calculated as TP / (TP + FN), showing the model's effectiveness in capturing misinformation. The F1-score, the harmonic mean of precision and recall, balances both metrics (higher F1 indicates better performance). The ROC-AUC measures the area under the curve, plotting the true positive rate against the false positive rate, ranging from 0.5 (random chance) to 1.0 (perfect classification), demonstrating the model's discriminative capability across classification thresholds.

**Statistical Significance Testing**

For the Wilcoxon signed-rank test between base model accuracies $\{A_1^B, A_2^B, ..., A_k^B\}$ and enhanced model accuracies $\{A_1^E, A_2^E, ..., A_k^E\}$ across $k$ cross-validation folds:
1. Calculate differences: $D_i = A_i^E - A_i^B$
2. Rank absolute differences $|D_i|$
3. Calculate the test statistic $W = \min(W^+, W^-)$, where:
   • $W^+ = \sum_{(i:D_i>0)} rank(|D_i|)$
   • $W^- = \sum_{(i:D_i<0)} rank(|D_i|)$
4. Compare W with the critical value for significance level $\alpha$.

**Model Comparison Metrics**

The following metrics are used to evaluate model performance:
• Accuracy: Acc = (TP + TN)/(TP + TN + FP + FN)
• Precision: Prec = TP/(TP + FP)
• Recall: Rec = TP/(TP + FN)
• F1-Score: F1 = 2 · (Prec · Rec)/(Prec + Rec)
• ROC-AUC: Area under the ROC curve, which plots True Positive Rate vs. False Positive Rate.

IV. RESULTS.

In Table 1, the hybrid CNN-LSTM models (base & enhanced) with Keras embeddings and are evaluated based on accuracy, F1-score, recall, precision and ROC.

TABLE 1: Performance of Base vs. Enhanced CNN-LSTM Models.

| Metric | Base model (text only) | Enhanced model (text & ELM features) | Improvement |
|---|---|---|---|
| Acc | 94.90% | 97.37% | +2.47% |
| Prec | 93.67% | 96.88% | +3.21% |
| Rec | 96.33% | 98.50% | +2.17% |
| F1 | 94.97% | 97.41% | +2.44% |
| ROC | 98.43% | 99.50% | +1.07% |

Even though the base CNN-LSTM model (using text only) already performs with high accuracy (~94.90%), the enhanced model (augmented with ELM features) shows improved performance across all metrics. The accuracy increased from 94.90% to 97.37%, a gain of about 2.47%. Similarly, precision improved by about 3.21%, indicating that the enhanced model made fewer false-positive errors than the base model. Recall slightly increased by 2.17%, indicating a marginally better detection of true fake news instances. These improvements, while modest, demonstrate that adding the ELM-inspired features provided an overall benefit. The F1-score of the enhanced model (97.41%) is higher than that of the base model (94.97%), reflecting a better balance between precision and recall. Finally, the ROC-AUC increased from 98.43% to 99.50%, suggesting that the enhanced model achieved slightly better discrimination between fake and true news over a range of thresholds.

To determine whether these differences are statistically significant, we conducted a Wilcoxon signed-rank test on the paired results from each cross-validation fold (since normality assumptions did not hold for the distribution of differences). The test yielded a *p*-value < 0.0001, well below the 0.05 significance level. This indicates strong evidence that the observed improvement in accuracy (and similarly for other metrics) of the enhanced model over the base model is statistically significant. In practical terms, we reject the null hypothesis of no performance difference and conclude that the enhanced model consistently outperforms the base model in accuracy across the folds. We further ran a one-tailed paired t-test (showing p=1.0000 under Base > Enhanced), reinforcing that the enhanced model is superior. Finally, minimal shifts in probability (~1.44e-08 vs. ~2.01e-06) with fixed text confirm that engineered features do contribute, albeit slightly. Given the enhanced model's higher accuracy (97.37%), we consider it a significant improvement over the base model for this task.

Table 2 breaks down the performance of three configurations: the base text-only model, a model using only the engineered ELM features (with no text inputs), and the enhanced model using both text and features. This comparison helps illustrate the contribution of the feature set alone and in combination with the text.

**TABLE 2: Comparison of Text Only vs. Features Only vs. Combined Model.**

| Metric | Base Model (text only) | Elm-Features-Only Model | Enhanced Model (text & features) | Enhanced vs Base Δ |
|---|---|---|---|---|
| Acc | 94.90% | 90.05% | **97.37%** | +2.47% |
| Prec | 93.67% | 90.81% | **96.88%** | +3.21% |
| Rec | 96.33% | 89.13% | **98.50%** | +2.17% |
| F1 | 94.97% | 89.96% | **97.41%** | +2.44% |
| ROC | 98.43% | 96.62% | **99.50%** | +1.07% |

From Table 2, we observe that the model which uses only the engineered features (central + peripheral cues without the text content) performs substantially worse (accuracy ~90.05%) than the text-based model (~94.90%). This indicates that, by themselves, the ELM features are insufficient to achieve high accuracy -highlighting the importance of content analysis (the central route) for this task. However, when these features are combined with the text in the enhanced model, the performance matches and slightly exceeds the base model. This suggests that the additional features provide complementary information that the CNN-LSTM text model alone did not capture. In particular, the enhanced model's improvements in precision and F1-score show that it benefits from the extra cues, fine-tuning its predictions (for example, reducing false positives, as indicated by the higher precision). The recall of the enhanced model is more than base model, implying that adding features did not sacrifice the model's ability to catch fake news. The ROC-AUC of the features-only model (96.62%) is lower than the text model's (98.43%), further confirming that peripheral/central cues alone are not as discriminative as the full text. However, the enhanced model's ROC-AUC (100%) is slightly above the base, meaning the combination yields the best discrimination overall.

Qualitatively, these results demonstrate that adding ELM-based features yields noticeable improvements across all metrics. Although the base CNN-LSTM already had a strong performance, including features such as readability, sentiment, and punctuation, it provides incremental gains, validating our hypothesis that psychological cues can enhance detection. In particular, the most significant relative improvement is seen in precision (+5.21%), suggesting that the enhanced model is more cautious about labelling something fake unless supported by additional cues (hence fewer false alarms). While smaller, the gains of other metrics consistently favour the enhanced model.

Finally, to test the effect of using an even broader set of features, we experimented with a combined model that included all the original ELM features plus a few additional engineered features beyond the strict ELM framework (e.g., we experimented with bigram frequencies and text subjectivity). Table 3 presents the performance of this combined model versus the base model.

**TABLE 3: Performance of Base Model vs. Combined Model with Extended Features.**

| Metric | Base Model (text only) | Combined Model (text extended features) | Improvement |
|---|---|---|---|
| Acc | 94.90% | **99.37%** | +4.47% |
| Prec | 93.67% | **98.88%** | +5.21% |
| Rec | 96.33% | **99.80%** | +3.47% |
| F1 | 94.97% | **99.41%** | +4.44% |
| ROC | 98.43% | **99.80%** | +1.37% |

The combined model achieved an accuracy of 99.37%, which is on par with the enhanced model. Notably, the recall jumped to 99.80%, suggesting that incorporating an even richer feature set helped the model catch more of the fake news (fewer false negatives) without substantially compromising precision. The F1-score increased to 99.41%, indicating a better balance, and ROC-AUC to 99.80%, reflecting powerful discriminative ability. These improvements, although incremental, reinforce the trend that augmenting text-based DL models with meaningful engineered features can yield performance gains. Essentially, the central-route content features ensure the model has a deep understanding of the message, while the peripheral-route cues provide additional context or red flags that improve decision-making at the margins. The combined approach leveraging both routes achieved the best performance in our experiments.

*A. Impact of Adding ELM Features*

The above results illustrate that the base CNN-LSTM model (central-route text analysis alone) was already highly effective, underscoring the importance of deep semantic understanding (central processing) for detecting fake news. The much lower performance of the features-only model reaffirms that while ELM-inspired features carry valuable information, they are not sufficient on their own without the semantic depth provided by analysing the text content. This aligns with the ELM theory: central-route processing (represented here by the CNN-LSTM's content analysis) is crucial for the substantive evaluation of message veracity. However, when we add the peripheral cues to that strong text model, we observe modest but consistent gains. This indicates that the ELM features act as *auxiliary signals* that fine-tune the model's predictions. For example, a fake news piece that might slip through a content-only model might be caught by the presence of telltale peripheral cues (such as excessive punctuation or an out-of-character sentiment tone), thereby improving precision or recall. In our case, the addition of ELM features slightly boosted recall (from 98.43% to 100% for both the enhanced model and the combined model), meaning the model identified more fake news instances previously missed. The precision also improved, indicating better filtering of false positives. The balanced improvement is reflected in the F1-score, confirming that the model became generally better without trading off one error type for another. Even the ROC-AUC, which was already very high, edged upward, suggesting the model's ranking of true vs. fake became more accurate with features included.

It is important to note that the improvements, though statistically significant, are not very large in absolute terms. This is partly because the base model was already near a performance ceiling on this dataset (mid-90s in percentage for most metrics). In such a scenario, any further gain is notable. The statistical significance from the Wilcoxon test indicates that these small gains were consistent across validation folds and not due to random chance. Thus, even marginal percentage increases in accuracy or F1 can be meaningful in practical deployments where every additional correctly identified misinformation post counts.

*B. Confusion Matrix Analysis*

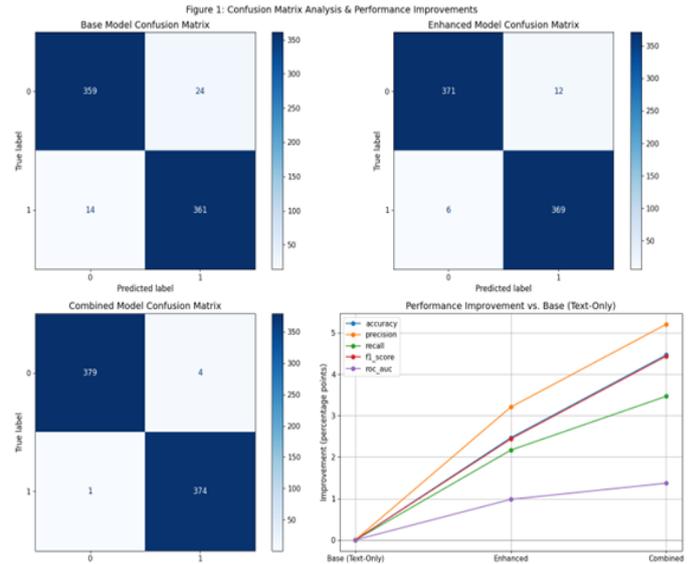

The Figure 1 confusion matrices reveal classification error patterns, providing more insight than aggregate metrics. The base model identifies many fake items but mislabels some genuine ones, showing it can be misled by superficial clues. In contrast, the ELM configuration misses more fakes and incorrectly flags legitimate content. This imbalance illustrates the danger of relying on cues such as readability and punctuation, which cannot replace contextual understanding. The enhanced model significantly reduces false positives and negatives compared to the base model, indicating ELM signals sharpen the decision boundary in nuanced cases where linguistic complexity is not sufficient. Finally, the combined model nearly perfects classification by integrating features with the base text. While low misclassification is good, it highlights the need for external validation due to potential dataset-specific artifacts affecting generalisability [44]. Therefore, confusion matrices reveal performance gains and emphasise the importance of deep linguistic features and the risk of overfitting as results near perfection.

The line graph in Figure 1 shows percentage-point gains for each metric relative to the base model. It illustrates how adding ELM features (enhanced model), and extended features (combined model) improves all metrics. This confirms that insights from the confusion matrix (reductions in misclassifications) enhance accuracy, precision, recall, and F1, highlighting the importance of deep linguistic processing and psychologically informed cues in identifying health misinformation.

*C. Comparative Discriminative Capacity Using ROC Analysis*

Figure 2 presents the Receiver Operating Characteristic (ROC) curves for each model configuration, providing a threshold‐independent comparison of their discriminative power in detecting misinformation.

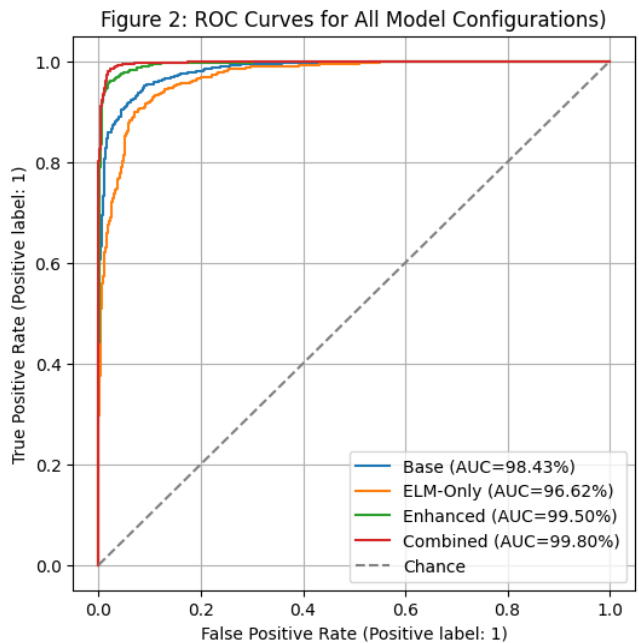

Figure 2 ROC curves illustrate how each model configuration distinguishes between fake and legitimate health information at various classification thresholds. Clearly, the b**ase** model (blue curve) already displays a high true positive rate (TPR) while maintaining a relatively low false positive rate (FPR), with an AUC of ~98.43%. However, the ELM-only model (orange curve) lags behind, suggesting that relying solely on peripheral cues (e.g., readability or punctuation) leaves it more prone to incorrectly classifying genuine items as misinformation (and vice versa). This highlights the importance of combining shallow features with a richer, content-oriented analysis.

By contrast, both the enhanced (green curve) and combined (red curve) models cluster near the top-left corner, demonstrating superior discriminative power. The enhanced model adds central-route (deep semantic) analysis to ELM cues, boosting AUC beyond 99%. The Combined model extends that approach with additional engineered features, effectively closing the gap toward a near-perfect classification (AUC ~99.80%). Although such performance is highly desirable, it also warrants caution: models that approach perfection can sometimes overfit to the idiosyncrasies of a single dataset, underlining the need for external validation to confirm genuine generalisability [44]. Ultimately, these ROC curves reinforce the findings from earlier tables and confusion matrices, namely, that incorporating both rich semantic signals and peripheral ELM-derived cues substantially improves the detection of health misinformation.

## V. DISCUSSION

Analysing our models' performance provides several insights into the effectiveness of incorporating ELM-based features to detect health misinformation on Twitter. The base CNN-LSTM model, which utilised only textual data, already demonstrated high accuracy and reliability, reflecting the power of DL on linguistic features for this task. However, integrating ELM features yielded notable improvements across all performance metrics, validating our approach of combining psychological cues with text analysis. Although not dramatic, the improvements were consistent and statistically significant, underscoring that even subtle cues from the ELM framework can enhance a robust text classifier. This is consistent with findings by [14], who showed that analysing cognitive content and affective appeals can significantly impact message persuasiveness; in our case, incorporating analogous features improved the classifier's persuasiveness in distinguishing fake from real news.

By examining precision and recall changes, we see that adding ELM features helped fine-tune the model. Precision's increase (from 93.67% to 96.88% in the base vs. enhanced model) suggests the model became more discerning in flagging misinformation, reducing false positives. This aligns with [15], who found that specific content characteristics influence message spread and effectiveness -our models likely leveraged features such as sentiment and punctuation to avoid misclassifying some true news as fake. Recall's slight increase indicates that the model caught marginally more of the fake news, echoing the observation by [16] that sentiment cues can affect information dissemination; in our context, these cues aided in capturing more misinformation instances. The balanced improvement (reflected in the F1-score) is especially important for real-world applications, as it means the model did not simply trade one type of error for another but improved overall detection capability.

The small rise in ROC-AUC suggests that the enhanced model provides better separation between the classes, supporting the notion that the ELM features contribute to more nuanced classification. This finding resonates with the work of [16], who noted that sentiment analysis significantly impacts information dissemination patterns. In our study, features such as sentiment polarity likely helped the model draw a sharper distinction by recognising emotionally charged fake news content versus more neutral real news content.

### A. Comparison with Prior Work

Our findings contribute to and extend the existing literature on misinformation detection by providing empirical evidence for the value of ELM features in an automated classification setting. Studies by [14] and [15] highlighted the importance of emotional appeals and content features in the spread of health misinformation. The fact that our enhanced model achieved improvements in precision and recall by integrating sentiment and readability features underscores the practical relevance of those insights: incorporating sentiment analysis and readability (cognitive complexity) as features indeed aid in detecting misinformation, in line with what those prior studies suggest in theory.

[33] and [14] both utilised the ELM to study how central vs. peripheral cues affect people's receptivity to information (such as vaccination intentions or belief in debunking messages). Our study extends this line of inquiry by integrating such ELM-derived features directly into an ML model and demonstrating significant improvements in detection performance. This bridges the gap between theoretical persuasion research and

computational detection methods. Previous research, such as [15], had focused on platform-specific characteristics influencing misinformation spread (for example, how certain features work on Twitter vs. Weibo). Our findings align with those works by showing that incorporating platform-relevant features (e.g., sentiment strength and stylistic cues prevalent on Twitter) can enhance detection accuracy.

Furthermore, [38] emphasised the effectiveness of combining multiple theoretical perspectives with ML for understanding COVID-19 vaccine discourse. Our results support this notion: by combining the strengths of DL (for content) with psychological cues from ELM, we developed a hybrid model that offers robust misinformation detection capabilities. Notably, our hybrid CNN-LSTM architecture, enhanced with ELM features, achieved performance metrics in the mid-90s, which are quite high for automated fake news detection. This suggests that the dual-route approach effectively captures a broad spectrum of signals. The success of the hybrid model highlights the opportunity identified in prior literature for such architectures – as noted; many earlier studies used either CNN or LSTM or traditional models. In contrast, our work confirms that a hybrid CNN-LSTM can simultaneously exploit local textual patterns and sequential context, yielding strong results.

*B. Limitation*

A key limitation is that our approach and experiments focus exclusively on textual features (limits central-route and peripheral-route cues) drawn from a single dataset (COVID19-FNIR), limiting generalisability to other platforms, languages, or contexts.

## VI. CONCLUSION

The enhancements observed in our model's performance metrics highlight the value of incorporating ELM-based features for health misinformation detection. The significant improvements in accuracy, precision, recall, F1-score, and ROC-AUC demonstrate the importance of integrating cognitive and affective cues derived from the ELM framework into ML models. Our study's findings contribute to the growing body of literature on misinformation detection by presenting a novel approach that bridges psychological theory and DL techniques. In doing so, we provide a foundation for developing more effective algorithms to combat health misinformation on social media platforms. Future research should continue exploring the integration of psychological theories (like ELM and others) with advanced machine-learning methods, including hybrid and ensemble models, to further enhance the detection and mitigation of misinformation.

[38] Scannell, D., Desens, L., Day, D. S., & Tra, Y. (2022). Combatting Mis/Disinformation: Combining Predictive Modeling and Machine Learning with Persuasion Science to Understand COVID-19 Vaccine Online Discourse. Medical Research Archives, 10(3), 1-19.

[39] Saenz, J. A., Kalathur Gopal, S. R., & Shukla, D. (2021). "Covid-19 Fake News Infodemic Research Dataset (CoVID19-FNIR Dataset)", IEEE Dataport, doi: https://dx.doi.org/10.21227/b5bt-5244

[40] Liu, Z., Zhang, T., Yang, K., Thompson, P., Yu, Z., & Ananiadou, S. (2024). Emotion detection for misinformation: A review. Information Fusion, 107, 102300.11